\documentclass[%
 reprint,
 amsmath,amssymb,
 aps,
]{revtex4-2}

\usepackage{color}
\usepackage{graphicx}
\usepackage{dcolumn}
\usepackage{bm}

\newcommand{\beq}{\begin{equation}}
\newcommand{\eeq}{\end{equation}}

\def\cE{ \mathcal{E} }
\def\cH{ \mathcal{H} }
\def\cI{ \mathcal{I}_\cE }

\newcommand{\bu}{{\bf u}} 
 
\newcommand{\bF}{{\bf f}}
\newcommand{\bk}{{\bf k}}

\newcommand{\bb}{{\bf b}}

\newcommand{\curl}{\nabla \times}

\usepackage[dvipsnames]{xcolor}
\def\NEW#1{{\textcolor{black}{#1}}}

\begin{document}

\preprint{APS/123-QED}


\title{Saturation of turbulent helical dynamos}

\author{Guillaume Bermudez}
\author{Alexandros Alexakis}%
 \email{alexakis@phys.ens.fr}
\affiliation{
Laboratoire de Physique de l’Ecole Normale Supérieure, ENS, Université PSL, CNRS, Sorbonne Université, Université Paris {~Cité}, F-75005 Paris, France
}%

\date{\today}

\begin{abstract}
The presence of large scale magnetic fields in nature is often attributed to the inverse cascade of magnetic helicity driven by turbulent helical dynamos. In this work we show that in turbulent helical dynamos, the inverse flux of magnetic helicity towards the large scales $\Pi_{\mathcal{H}}$ is bounded by $|\Pi_{\mathcal{H}}|\le  c \epsilon k_\eta^{-1}$, where  $\epsilon$ is the energy injection rate, $k_\eta$ is the Kolmogorov magnetic dissipation wavenumber \NEW{and $c$ an order one constant.} Assuming the classical isotropic
turbulence scaling,
the inverse flux of magnetic helicity $\Pi_\cH$ decreases at least as \NEW{a $-3/4$ power-law with the magnetic Reynolds number} $Rm$ : $|\Pi_\cH | \le c \epsilon \ell_f Rm^{-3/4}\max[Pm,1]^{1/4}$, where \NEW{$Pm$ the magnetic Prandtl number and $\ell_f$ the forcing lengthscale.}
We demonstrate this scaling with $Rm$ using direct numerical simulations of turbulent dynamos forced at intermediate scales. The results further indicate that nonlinear saturation is achieved by a balance between the inverse cascade and dissipation at domain size scales $L$
\NEW{for which} the saturation value of the magnetic energy
is bounded by $\cE_\text{m}\leq c L (\epsilon \ell_f)^{2/3} Rm^{1/4}\max[1,Pm]^{1/4}$. Numerical simulations also demonstrate this bound. 

\end{abstract}

\maketitle


\section{\label{sec:Intro} Introduction }  

Magnetic fields are observed in a plethora of astrophysical objects from planetary to galactic scales \cite{falgarone2008turbulence,ruzmaikin1988magnetic,parker1979cosmical}.  Their generation and sustainment is often attributed to dynamo action: their self-amplification by a continuous stretching and refolding of magnetic field lines due to the underlying (turbulent in most cases) flow \cite{moffatt1978field,rincon2019dynamo}.  In many cases the magnetic structures formed span the entire astrophysical object, reaching scales much larger than the small scale turbulence that generates them. 
The pioneering work of \cite{steenbeck1966berechnung} showed that large scale magnetic fields can be generated from small scale flows if the advecting flow is helical. 
This result is based on an expansion for large scale separation and is referred as alpha-dynamo. 
However, such expansion can be formally done only below a critical value of \NEW{the magnetic Reynolds number}
 $Rm$ (the ratio of Ohmic to dynamic time scales).
Above this critical value 
small scale dynamo action begins and the expansion ceases to be valid \cite{boldyrev2005magnetic, 
cattaneo2009problems, 
cameron2016fate}. 

The validity of the alpha model is further questioned in the nonlinear regime for which the magnetic field feeds back to the velocity field through the Lorentz force. In~\cite{vainshtein1992nonlinear,hughes2008mean} it was argued that the growth of alpha dynamos saturates when the large scale magnetic field $\bf B$ becomes larger than $U_\text{rms} Rm^{-1/2}$, where $U_\text{rms}$ is the root mean square value of the velocity fluctuations. This gives a very weak magnetic field for most astrophysical applications for which $Rm\gg 1$. 
\NEW{ Two scale models have also been extensively used
to predict saturation magnetic energy 
\cite{blackman2002new,blackman2003recent} 
but their application is limited for large $Rm$ where turbulence sets in. }
%

An alternative way of explaining the formation of large magnetic fields is through the inverse cascade
of magnetic helicity~\cite{frisch1975possibility,pouquet1976strong}. This  intrinsically nonlinear mechanism (that is however compatible with alpha dynamos)  predicts that magnetic helicity will be transferred by nonlinear interactions to larger scales. Indeed, several works that followed 
\cite{pouquet1978numerical,brandenburg2001inverse,alexakis2006inverse,muller2012inverse,bhat2019efficient}
demonstrated with numerical simulations that when magnetic helicity is injected in a flow (by a dynamo or other mechanism), it cascades inversely to larger scales.
These results however were performed on moderate values of $Rm$ 
\NEW{(a few times the small scale dynamo onset $Rm_\text{c} \simeq10$)}
and the inverse cascade of magnetic helicity at large $Rm$ is not tested. To our knowledge a quantitative understanding of how turbulent helical dynamos saturate 
with clear predictions of the saturating amplitude does not exist.

In this work we demonstrate by analytical arguments and numerical simulations that in dynamo flows, the inverse cascade of magnetic helicity is bounded from above by a decreasing power of $Rm$. Thus it cannot survive the infinite $Rm$ limit.
This leads to a prediction for the saturation amplitude of the magnetic field that we test with numerical simulations.

\section{\label{sec:Theor} Theoretical arguments} 
We begin by considering the  MHD equations 
for the incompressible velocity $\bf u$ and magnetic field $\bf b$ given by
\begin{eqnarray}
\partial_t {\bf u}+ {\bf u}\cdot {\bf \nabla u} &=&
-{\bf \nabla} P + {\bf (\nabla \times b)\times b} 
+\nu \nabla^2 {\bf u} + {\bf f}, \label{mhd1}\\
\partial_t {\bf b} &=&
 {\bf \nabla \times (u\times b)} 
+\eta \nabla^2 {\bf b},
\label{mhd2}
\end{eqnarray}
in a cubic periodic domain of size $2\pi L$,
with $\nu$ being the viscosity, $\eta$ the magnetic diffusivity, \NEW{$P$ the pressure,} and $\bf f$ an external mechanical force.
\NEW{In absence of viscous and Ohmic dissipation, and external forcing,} the total energy $\mathcal{E}=\frac{1}{2}\langle |{\bf u}|^2+|{\bf b}|^2\rangle$ and magnetic helicity 
$\mathcal{H}=\frac{1}{2} \langle {\bf a\cdot b} \rangle$
\NEW{are conserved ; here the angular brackets stand for spatial integration}
and ${\bf a}= -\nabla^{-2}\nabla\times {\bf b}$ is the vector potential. 
Their balance reads
\beq
\partial_t  \cE = \cI - \epsilon, \qquad \partial_t  \cH = -\epsilon_{_\cH},
\eeq 
where $\cI=\langle \bF\cdot \bu \rangle$ is the energy injection rate, $\epsilon=\epsilon_u+\epsilon_b$ is the energy dissipation rate with $\epsilon_u=\nu \langle |\nabla \bu |^2 \rangle $ the viscous dissipation rate and $\epsilon_b=\eta \langle |\nabla \bb |^2 \rangle $ the Ohmic dissipation rate.
Finally $\epsilon_{_\cH}=\eta \langle \bb \cdot \curl \bb \rangle$ is the helicity generation/dissipation rate.
The forcing $\bf f$ is assumed to act on a scale 
\NEW{$\ell_f=k_f^{-1}\ll L$ }
while dissipation acts 
at the smaller viscous scale $\ell_\nu$ and Ohmic scale $\ell_\eta$.
For large Reynolds number $Re=\epsilon^{1/3}\ell_f^{4/3}/\nu$ and large magnetic Reynolds number
$Rm=\epsilon^{1/3}\ell_f^{4/3}/\eta$, 
the dissipation length scales $\ell_\nu,\ell_\eta$
%
scale like \cite{schekochihin2002spectra}
\beq 
\ell_\nu \propto \ell_f Re^{-3/4} \quad \mathrm{and} \quad \ell_\eta \propto \ell_f Rm^{-3/4} 
\eeq 
for $Pm\le 1$ while for $Pm\ge 1$ 
\beq 
\ell_\nu \propto \ell_f Re^{-3/4} \quad \mathrm{and} \quad  
\ell_\eta \propto 
                   \ell_f Rm^{-3/4}Pm^{1/4},
\eeq 
\NEW{where $Pm = \nu/\eta$ is the magnetic Prandtl number.}

At intermediate scales $\ell$ (so called inertial scales $\ell_f \ll \ell \ll \ell_\eta,\ell_\nu$), there is a constant flux of energy across scales given by
\beq 
\Pi_\cE(k)=\langle \bu_k^< \cdot ( \bu \cdot \nabla \bu - \bb \cdot \nabla \bb ) - \bb_k^< \cdot (\nabla \times \bu \times \bb)   \rangle, 
\eeq 
where $ \bu_k^<,\bb_k^<$ stand for the filtered velocity and magnetic field respectively 
so that only Fourier modes with wavenumbers of norm smaller than $k=1/\ell$ are kept \cite{biskamp2003magnetohydrodynamic}.
Conservation of energy by the nonlinear terms implies that the
energy flux at the inertial scales is constant in $k$ and equals the energy dissipation rate $\Pi_\cE(k)=\epsilon$. 

Similarly, there is a flux of magnetic helicity \NEW{\cite{alexakis2006inverse}}
\beq 
\Pi_\cH(k)=- \langle \bf b_k^< \cdot  ( u \times  b )   \rangle 
\eeq
that also has to be constant at scales that dissipation plays no role. However, unlike energy,
the forcing  does not inject magnetic helicity which can be only generated or destroyed by the Ohmic dissipation
at  rate $\epsilon_{_\cH}$. Nonetheless if the forcing is helical the flow can transport magnetic helicity from the small Ohmic scales to ever larger scale $L'>\ell_f$
up until the domain size reached $L'\simeq L$ where a helical condensate will form. Conservation of magnetic helicity by the nonlinear terms implies again that the
flux of helicity at scales $L'\gg \ell \gg \ell_\eta$ has to be constant in $k\propto1/\ell$ with  $\Pi_\cH(k)=\epsilon_{_\cH}$.

The two cascades, energy and helicity, are not independent and the first limits the later \cite{alexakis2018cascades}.  
To show that, we write the magnetic field in Fourier space 
$\tilde{\bf b}_{\bf k}$ using the helical basis $\tilde{\bf b}_{\bf k}=b_{\bf k}^+{\bf h^+_k}+b_{\bf k}^-{\bf h^-_k}$ where 
\beq 
{\bf h_k^\pm}= \frac{\bf e\times k \times k}{\sqrt{2}|{\bf e\times k \times k}|} \pm
          i  \frac{\bf e \times k}{\sqrt{2}|{\bf e\times k }|}
\eeq 
are the eigenvectors of the curl operator with $\bf e$ an arbitrary vector \NEW{(non-parallel to $\bf k$)}
\cite{waleffe1992nature,linkmann2016helical}. By doing that we can write the magnetic energy spectrum as $E_b(k)=E_b^+(k)+E_b^-(k)$
and the magnetic helicity spectrum as $H(k)=(E_b^+(k)-E_b^-(k))/k$ where 
$E_b^\pm(k) \text{d}k = \sum_{k\le|\bk|<k+\text{d}k} |b_{\bf k}^\pm|^2 $ is the sum of the energy of the $b^\pm$
Fourier modes on a spherical shell of width $\text{d}k$ and radius $k$. 
%
Since magnetic helicity is primarily generated at Ohmic 
wavenumbers $k_\eta=1/\ell_\eta$
for any wavenumber $k$ in the range $k\le k_\eta$ we can write 
\begin{eqnarray}
|\Pi_\cH(k)| &=& | \epsilon_{_\mathcal{H}} | \nonumber \\
 &\simeq& \eta \left| \int_k^\infty 
q[E_b^+(q)-E_b^-(q)]\text{d}q\right| \nonumber \\
&\le& \eta k^{-1}  \int_k^\infty q^2 \left|E_b^+(q)+E_b^-(q) \right| \text{d}q \nonumber \\
&=&  k^{-1} \epsilon_b  
\label{MS}
\end{eqnarray}
\NEW{This result holds for any $k$ in the inertial range.
Choosing $k=k_\eta/c$, where $c>1$ is an order one constant,  
and using $\epsilon_b \le \epsilon $, }
we obtain our final result
\beq 
|\Pi_\cH(k)| \le c \, \epsilon /k_\eta.
\label{MR1}
\eeq 
\NEW{In other words the maximum possible value of $|\epsilon_{_\mathcal{H}}|$ is obtained if all magnetic energy at Ohmic scales $k_\eta^{-1}$ is concentrated at only positive or only negative helicity modes, in which case
$|\epsilon_{_\mathcal{H}}|=\epsilon_b/k_\eta$.
Thus the flux of magnetic helicity is bounded by the energy injection rate divided by the Ohmic dissipation wavenumber $k_\eta$}. 
Note that this bound is saturated if the magnetic field at small scales is fully helical.
If not, $|\Pi_\cH|$ can be much smaller than \eqref{MR1}. 
Using the estimates for $k_\eta$ for isotropic MHD turbulence we obtain
\beq |\Pi_\cH(k)| \le c \epsilon k_f^{-1} Rm^{-3/4} \max[1,Pm]^{1/4}, \label{MR2}  \eeq
where $c$ is an order one constant.
Given the very large values of $Rm$ in nature this gives very little hope of observing such fluxes. 

However, despite having a diminishing flux of magnetic helicity for large $Rm$, this does not mean that large scale dynamos cannot be observed. In a finite domain of size $L$ a magnetic helicity condensate will form. Its magnetic field amplitude $B$ will be determined by a balance of magnetic helicity flux with the magnetic helicity dissipation at that scale so that  
$\Pi_\mathcal{H} \propto  \eta B^2/L $. Using the previous estimate for the flux
leads to the prediction for the large scale energy $\cE_m=B^2$ given by  
\beq \cE_m \le \NEW{c\,} \epsilon^{2/3} L k_f^{1/3} Rm^{1/4}\max[1,Pm]^{1/4}. \label{MR3} \eeq
%
We note that a very long time $T\sim (B^2L)/\Pi_\mathcal{H}\propto Rm $ would be required
for such a field to be formed.
Furthermore if  there is an other $\eta$-independent mechanism for magnetic helicity saturation, 
like magnetic helicity expulsion \cite{rincon2021helical,cattaneo2020magnetic},
then the amplitude of the large scale magnetic field will diminish to zero as $Rm\to\infty.$ 
 
Finally we note that this result is based on strong turbulence scaling of $k_\eta$.
One can argue that as the large scale magnetic field builds up the relation
between $k_\eta$ and $Rm$ can change from that of strong turbulence to that of 
weak turbulence \cite{galtier2000weak} or turbulence driven by the large scale magnetic shear \cite{alexakis2013large}. 
Both of these options lead to a faster increase of $\mathcal{E}_m$ with $Rm$ as  $\cE_m \lesssim \NEW{c\,} \epsilon^{2/3} L^{5/6} k_f^{-8/15} Rm^{2/5} $
that is not found however in the numerical studies that follow.
%
\section{\label{sec:Numer} Numerical simulations} 

To demonstrate the above arguments we perform a series of numerical simulations that solve the MHD equations (\ref{mhd1}--\ref{mhd2}) using the pseudo-spectral code {\sc Ghost} \cite{mininni2011hybrid} in a cubic domain of side $2\pi L$ with a fully helical random delta correlated forcing at wavenumber $k_f=1/\ell_f$ that fixes the energy injection rate $\epsilon$. The \NEW{magnetic} Prandtl number $Pm=\nu/\eta$ was set to unity for all runs. The resolution used varied from $128^3$ grid points to $1024^3$ grid points for the largest $Rm$.
The resolution was chosen so that the largest inertial range is obtained while remaining well resolved with a clear dissipation wavenumber range.
\NEW{All measurements were obtained by time averaging
at steady state. }

As a first step to accommodate for the the large scale pile up of magnetic helicity we introduce a magnetic hypo-dissipation term $\eta_h \nabla^{-2} {\bf b}$ in \eqref{mhd2} that arrests large scale magnetic helicity. With the inclusion of this term, the simulations reach quickly a steady state
where the helicity generated at the small scales by $\epsilon_\cH$ is transported and dissipated at the largest scales by hypo-dissipation.
\begin{figure}
    \centering
    \includegraphics[width=0.5\textwidth]{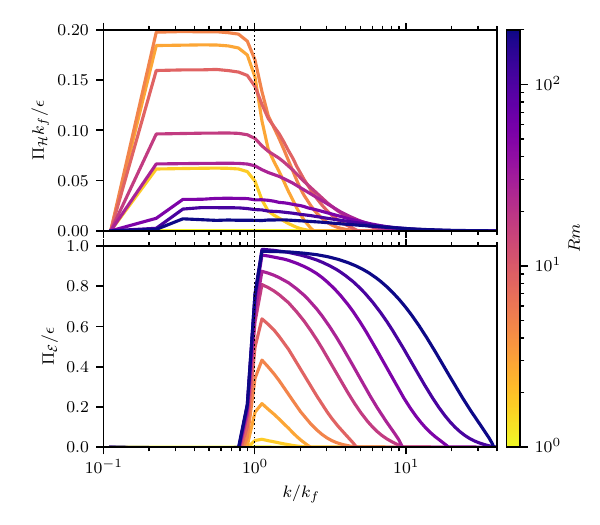}\vspace{-2em}
    \caption{Magnetic helicity flux $\Pi_\cH$ (top) and total energy flux $\Pi_\cE$ (bottom) for several values of the magnetic Reynolds number $Rm$ for $k_fL = 8$.}
    \label{fig:flux}
\end{figure}
\begin{figure}
    \centering
    \includegraphics[width=0.5\textwidth]{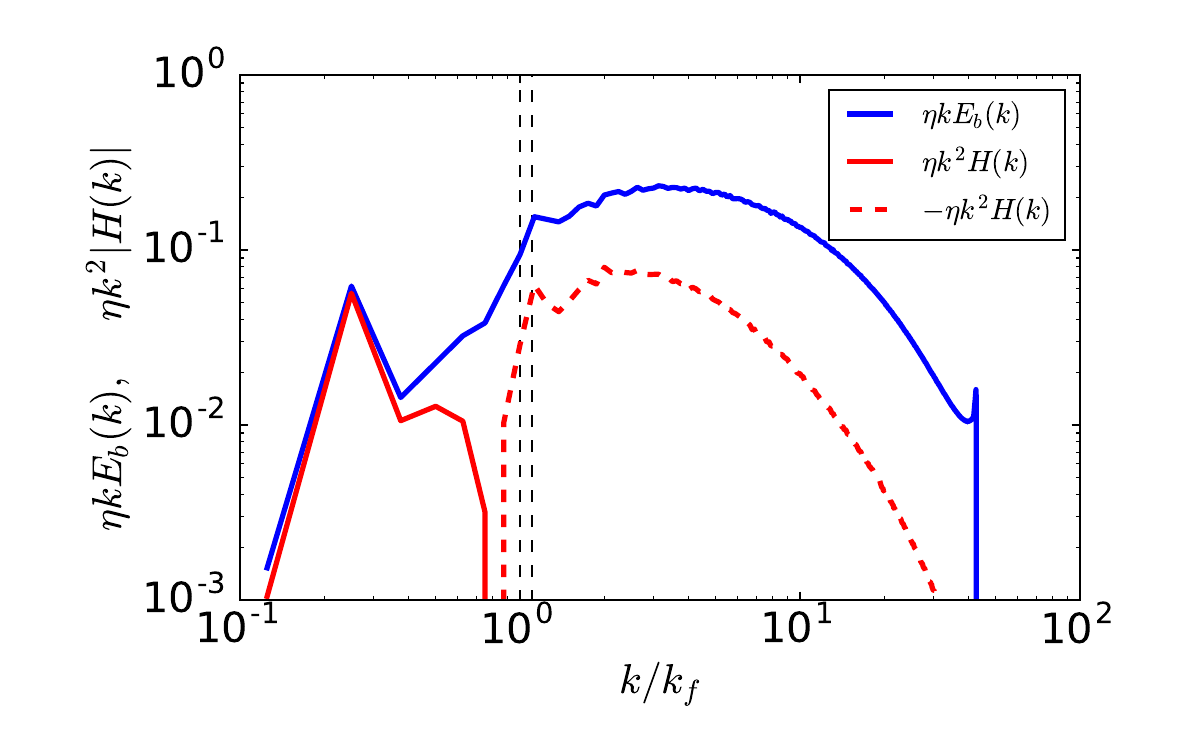}\vspace{-2em}
    \caption{ 
    Magnetic helicity dissipation/generation spectrum $\eta k^2 H(k)$ and normalized magnetic energy spectrum $\eta kE_b(k)$   for the case with hypo-dissipation with $k_f L= 8$ and the largest attained magnetic Reynolds number $Rm$.
    }
    \label{fig:Spec1}
\end{figure}

Figure \ref{fig:flux} shows the magnetic helicity flux (top) and energy flux (bottom) for a series of runs varying $Rm$ as shown in the legend for $k_fL=8$. 
The energy flux shows its classical behavior increasing as $Rm$ is increased, approaching its maximal value $\Pi_\cE\sim\epsilon$ in the inertial range. On the other hand,  $\Pi_\cH(k)$ first  starts to increase with $Rm$ once the dynamo onset is crossed
\NEW{(and magnetic energy and helicity appear in the system)}, reaches a maximum and then starts to decrease again
\NEW{when the cascades builds up so that the constrain in \eqref{MR1} becomes relevant. } 

The magnetic helicity generation spectrum $\eta k^2 H(k)$  for  the largest 
$Rm$ is shown in figure \ref{fig:Spec1}. 
Solid line corresponds to positive values of $\eta k^2H(k)$  while dashed line corresponds to negative values of $\eta k^2 H(k)$, thus negative magnetic helicity 
is generated at the smallest scales. In the same plot we show
the normalized magnetic energy spectrum $\eta kE_b(k)$ that bounds $\eta k^2|H(k)|\le \eta kE_b(k)$ 
with the equality corresponding to a fully helical magnetic field. 
\begin{figure}
    \centering
    \includegraphics[width=0.5\textwidth]{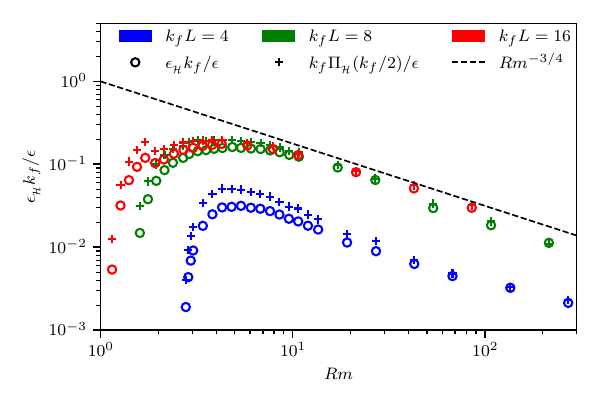}\vspace{-2em}
    \caption{Magnetic helicity generation rate $\epsilon_\cH$ (circles) and magnetic flux $\Pi_ \cH(k_f/2)$ (crosses) 
    as a function of the magnetic Reynolds number $Rm$ for three different values of $k_fL=4$ (blue),
    $k_fL=8$ (green) and $k_fL=16$ (red).}
    \label{fig:epsH}
\end{figure}

The dependence of the magnetic helicity cascade with $Rm$ is best seen in figure \ref{fig:epsH} where the magnetic helicity generation rate  $\epsilon_{_\mathcal{H}}k_f/\epsilon$ (circles) and the magnetic helicity flux $\Pi_\cH(k)$ at $k=k_f/2$ (crosses) are plotted as a function of $Rm$ for three different scale separations $k_fL=4,8,16$ in a log-log plot. All series show an initial increase of $\epsilon_{_\mathcal{H}}$ and $\Pi_\cH(k_f/2)$ followed afterwards by a power-law decrease
with $\epsilon_{_\mathcal{H}}\simeq\Pi_\cH(k_f/2)$. The dashed lines give the predicted scaling $Rm^{-3/4}$ that appears to fit very well the observed power verifying our prediction in \eqref{MR2}. 

\begin{figure}
    \centering
    \includegraphics[width=0.5\textwidth]{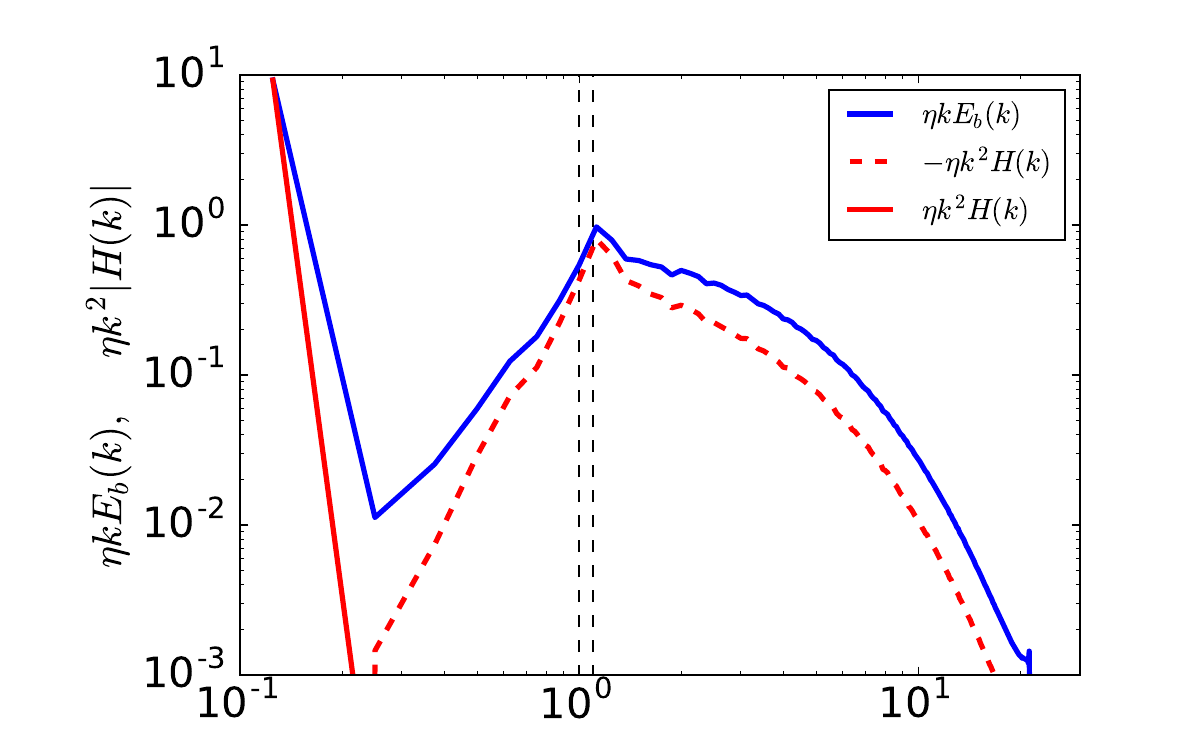}\vspace{-0.5em}
    \includegraphics[width=0.5\textwidth]{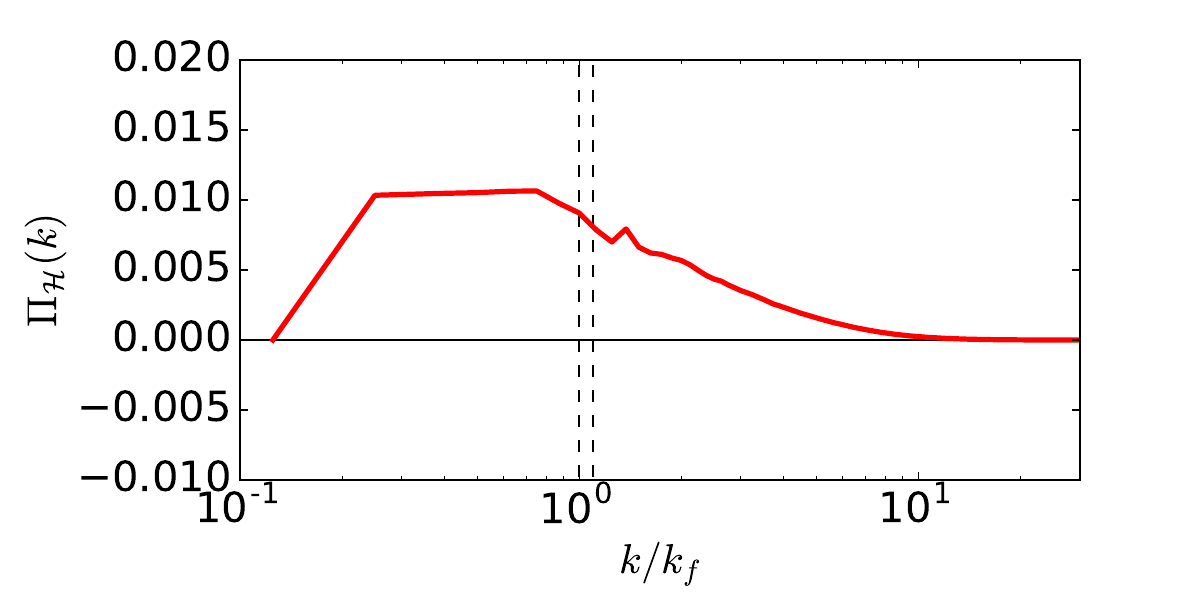}\vspace{-1em}
    \caption{Top panel: Magnetic helicity dissipation/generation spectrum $\eta k^2 H(k)$ and normalized magnetic energy spectrum $\eta kE_b(k)$ 
    for the runs with no hypo-dissipation, with $k_fL = 8$ and the largest attained magnetic Reynolds number $Rm$. Bottom panel: magnetic helicity flux for the same run.}
    \label{fig:Spec2}
\end{figure}

A second series of numerical simulations with $k_fL=8$ were performed without the hypo-viscous term.
The simulations were run for very long times until a steady state is reached such that magnetic energy 
does not increase further. The time scale to reach saturation is very long and this has limited us to use grids of size up to $512^3$
and values of $Rm$ four times smaller than the case with hypo-dissipation.
Figure \ref{fig:Spec2} shows with a red line the magnetic helicity dissipation spectrum $\eta k^2H(k)$ for the largest $Rm$ examined 
for these runs. Positive magnetic helicity is concentrated in a large scale condensate at $k=1/L=k_f/8$ while it is negative for all smaller scales. 
As in figure \ref{fig:Spec1} we also show $\eta k E_b(k)$ with a blue line.
Note that while the large scales are fully helical small scales are less. 
The amplitude of the large scale condensate (at the smallest $k=k_f/8$) is so large that despite the small value of $\eta$ the negative magnetic helicity generated at small scales 
by Ohmic diffusion is balanced by the positive helicity generated at the largest scale again by Ohmic diffusion. 
This leads to the flux of (negative) helicity from small to large scales shown in the
lower panel of the same figure.

\begin{figure}
    \centering
    \includegraphics[width=0.5\textwidth]{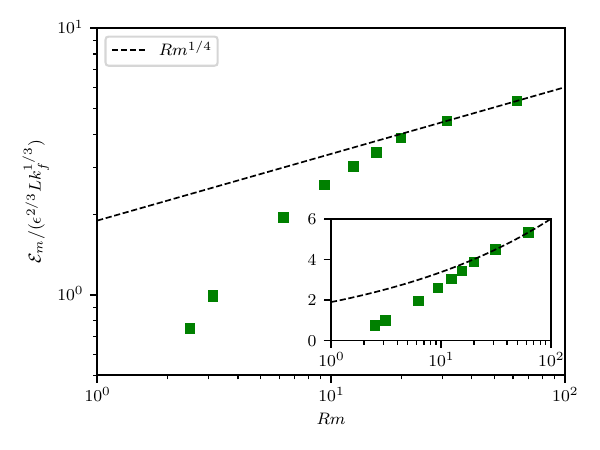}\vspace{-2em}
    \caption{Saturation of the magnetic energy $\cE_\text{m,sat}$ as a function of the magnetic Reynolds number $Rm$ for $k_fL = 8$
    in a log-log plot. The dashed line gives the prediction $Rm^{1/4}$. Inset: same figure in lin-log plot.}
    \label{fig:magnetic_field}
\end{figure}

The balance between  the magnetic helicity generated at small scales and dissipated at the 
large leads to the prediction \eqref{MR3} of a weak power-law increase of magnetic energy $\mathcal{E}_m$ with $Rm$.
Figure \ref{fig:magnetic_field} shows $\mathcal{E}_m$ as a function of $Rm$ in a log-log plot. 
The last three points appear to agree with the predicted power-law. 
The range  of values of $Rm$ compatible with this law is however rather limited and 
smaller power-laws or logarithmic increase cannot be excluded. In the inset of the same figure we show the same data
in a lin-log plot demonstrating that the data could also be fitted to a logarithmic increase. 
A logarithmic increase, if true, is still compatible
with the bound \eqref{MR3} but does not saturate it, and would imply that small scales become 
less and less helical as $Rm$ is increased. 

%
%
%



\section{\label{sec:Concl} Conclusion}

The present work gives for the first time estimates for the flux of magnetic helicity 
and the saturation of the magnetic field for the nonlinear state of a large $Rm$ turbulent helical dynamo.
Remarkably, even for asymptotically large values of $Rm$ it has been shown that
the magnetic helicity flux still depends on $Rm$ and in fact  it has to decrease at least as fast as 
$\Pi_{_\cH}\le  \epsilon/k_\eta \sim \epsilon/k_f Rm^{-3/4}$.
This analytical result has also been clearly demonstrated by numerical simulations
that are shown to follow this upper bound scaling. This excludes any inviscid nonlinear theory
for large scale dynamos from being realized!
%
 
Furthermore it was shown that
saturation is achieved by a balance of the inverse magnetic helicity flux with the helicity dissipation at the condensate scale. 
This has lead to the prediction that the magnetic field amplitude at steady state  is smaller than $Rm^{1/4}$. 
Numerical simulations are compatible with this result although the covered range of $Rm$ cannot exclude other small power-laws or logarithmic increase with $Rm$. 

It is important to note that these results are based on magnetic helicity conservation and are independent of the actual dynamo mechanism involved (alpha or other). They are thus rather general. 
Finally, we  stress that if  an $\eta-$independent mechanism exists at large scales to saturate the large scale magnetic helicity (like magnetic flux expulsion), our result \eqref{MR1} implies that no dominant large scale magnetic field will be present in the $Rm\to \infty$ limit.

The present results have critical implications for large magnetic fields in astrophysical systems and their origin.

\begin{acknowledgments}
%
This work was granted access to the HPC resources of MesoPSL financed by the Région Île-de-France and the project EquipMeso (project no. ANR-10-EQPX-29-01)
and of GENCI-TGCC \& GENCI-CINES (project no. A0090506421 and A0110506421).
This work been supported by the project Dysturb (project no. ANR-17-CE30-0004) finnanced by the Agence Nationale pour la Recherche (ANR).
\end{acknowledgments}



\bibliography{Magnetic_Helicity}

\providecommand{\noopsort}[1]{}\providecommand{\singleletter}[1]{#1}%
\begin{thebibliography}{30}%
\makeatletter
\providecommand \@ifxundefined [1]{%
 \@ifx{#1\undefined}
}%
\providecommand \@ifnum [1]{%
 \ifnum #1\expandafter \@firstoftwo
 \else \expandafter \@secondoftwo
 \fi
}%
\providecommand \@ifx [1]{%
 \ifx #1\expandafter \@firstoftwo
 \else \expandafter \@secondoftwo
 \fi
}%
\providecommand \natexlab [1]{#1}%
\providecommand \enquote  [1]{``#1''}%
\providecommand \bibnamefont  [1]{#1}%
\providecommand \bibfnamefont [1]{#1}%
\providecommand \citenamefont [1]{#1}%
\providecommand \href@noop [0]{\@secondoftwo}%
\providecommand \href [0]{\begingroup \@sanitize@url \@href}%
\providecommand \@href[1]{\@@startlink{#1}\@@href}%
\providecommand \@@href[1]{\endgroup#1\@@endlink}%
\providecommand \@sanitize@url [0]{\catcode `\\12\catcode `\$12\catcode
  `\&12\catcode `\#12\catcode `\^12\catcode `\_12\catcode `\%12\relax}%
\providecommand \@@startlink[1]{}%
\providecommand \@@endlink[0]{}%
\providecommand \url  [0]{\begingroup\@sanitize@url \@url }%
\providecommand \@url [1]{\endgroup\@href {#1}{\urlprefix }}%
\providecommand \urlprefix  [0]{URL }%
\providecommand \Eprint [0]{\href }%
\providecommand \doibase [0]{http://dx.doi.org/}%
\providecommand \selectlanguage [0]{\@gobble}%
\providecommand \bibinfo  [0]{\@secondoftwo}%
\providecommand \bibfield  [0]{\@secondoftwo}%
\providecommand \translation [1]{[#1]}%
\providecommand \BibitemOpen [0]{}%
\providecommand \bibitemStop [0]{}%
\providecommand \bibitemNoStop [0]{.\EOS\space}%
\providecommand \EOS [0]{\spacefactor3000\relax}%
\providecommand \BibitemShut  [1]{\csname bibitem#1\endcsname}%
\let\auto@bib@innerbib\@empty
\bibitem [{\citenamefont {Falgarone}\ and\ \citenamefont
  {Passot}(2008)}]{falgarone2008turbulence}%
  \BibitemOpen
  \bibfield  {author} {\bibinfo {author} {\bibfnamefont {E.}~\bibnamefont
  {Falgarone}}\ and\ \bibinfo {author} {\bibfnamefont {T.}~\bibnamefont
  {Passot}},\ }\href@noop {} {\emph {\bibinfo {title} {Turbulence and magnetic
  fields in astrophysics}}},\ Vol.\ \bibinfo {volume} {614}\ (\bibinfo
  {publisher} {Springer},\ \bibinfo {year} {2008})\BibitemShut {NoStop}%
\bibitem [{\citenamefont {Ruzmaikin}\ \emph {et~al.}(1988)\citenamefont
  {Ruzmaikin}, \citenamefont {Sokoloff},\ and\ \citenamefont
  {Shukurov}}]{ruzmaikin1988magnetic}%
  \BibitemOpen
  \bibfield  {author} {\bibinfo {author} {\bibfnamefont {A.}~\bibnamefont
  {Ruzmaikin}}, \bibinfo {author} {\bibfnamefont {D.}~\bibnamefont {Sokoloff}},
  \ and\ \bibinfo {author} {\bibfnamefont {A.}~\bibnamefont {Shukurov}},\
  }\href {https://books.google.fr/books?id=DgiChm5RNyAC} {\emph {\bibinfo
  {title} {Magnetic Fields of Galaxies}}},\ Astrophysics and Space Science
  Library\ (\bibinfo  {publisher} {Springer Netherlands},\ \bibinfo {year}
  {1988})\BibitemShut {NoStop}%
\bibitem [{\citenamefont {Parker}(1979)}]{parker1979cosmical}%
  \BibitemOpen
  \bibfield  {author} {\bibinfo {author} {\bibfnamefont {E.~N.}\ \bibnamefont
  {Parker}},\ }\href@noop {} {\emph {\bibinfo {title} {Cosmical magnetic
  fields: Their origin and their activity}}}\ (\bibinfo  {publisher} {Oxford
  university press},\ \bibinfo {year} {1979})\BibitemShut {NoStop}%
\bibitem [{\citenamefont {Moffatt}(1978)}]{moffatt1978field}%
  \BibitemOpen
  \bibfield  {author} {\bibinfo {author} {\bibfnamefont {H.~K.}\ \bibnamefont
  {Moffatt}},\ }\href@noop {} {\emph {\bibinfo {title} {Field generation in
  electrically conducting fluids}}},\ Vol.~\bibinfo {volume} {2}\ (\bibinfo
  {year} {1978})\BibitemShut {NoStop}%
\bibitem [{\citenamefont {Rincon}(2019)}]{rincon2019dynamo}%
  \BibitemOpen
  \bibfield  {author} {\bibinfo {author} {\bibfnamefont {F.}~\bibnamefont
  {Rincon}},\ }\href@noop {} {\bibfield  {journal} {\bibinfo  {journal}
  {Journal of Plasma Physics}\ }\textbf {\bibinfo {volume} {85}} (\bibinfo
  {year} {2019})}\BibitemShut {NoStop}%
\bibitem [{\citenamefont {Steenbeck}\ \emph {et~al.}(1966)\citenamefont
  {Steenbeck}, \citenamefont {Krause},\ and\ \citenamefont
  {R{\"a}dler}}]{steenbeck1966berechnung}%
  \BibitemOpen
  \bibfield  {author} {\bibinfo {author} {\bibfnamefont {M.}~\bibnamefont
  {Steenbeck}}, \bibinfo {author} {\bibfnamefont {F.}~\bibnamefont {Krause}}, \
  and\ \bibinfo {author} {\bibfnamefont {K.-H.}\ \bibnamefont {R{\"a}dler}},\
  }\href@noop {} {\bibfield  {journal} {\bibinfo  {journal} {Zeitschrift
  f{\"u}r Naturforschung A}\ }\textbf {\bibinfo {volume} {21}},\ \bibinfo
  {pages} {369} (\bibinfo {year} {1966})}\BibitemShut {NoStop}%
\bibitem [{\citenamefont {Boldyrev}\ \emph {et~al.}(2005)\citenamefont
  {Boldyrev}, \citenamefont {Cattaneo},\ and\ \citenamefont
  {Rosner}}]{boldyrev2005magnetic}%
  \BibitemOpen
  \bibfield  {author} {\bibinfo {author} {\bibfnamefont {S.}~\bibnamefont
  {Boldyrev}}, \bibinfo {author} {\bibfnamefont {F.}~\bibnamefont {Cattaneo}},
  \ and\ \bibinfo {author} {\bibfnamefont {R.}~\bibnamefont {Rosner}},\
  }\href@noop {} {\bibfield  {journal} {\bibinfo  {journal} {Physical review
  letters}\ }\textbf {\bibinfo {volume} {95}},\ \bibinfo {pages} {255001}
  (\bibinfo {year} {2005})}\BibitemShut {NoStop}%
\bibitem [{\citenamefont {Cattaneo}\ and\ \citenamefont
  {Hughes}(2009)}]{cattaneo2009problems}%
  \BibitemOpen
  \bibfield  {author} {\bibinfo {author} {\bibfnamefont {F.}~\bibnamefont
  {Cattaneo}}\ and\ \bibinfo {author} {\bibfnamefont {D.}~\bibnamefont
  {Hughes}},\ }\href@noop {} {\bibfield  {journal} {\bibinfo  {journal}
  {Monthly Notices of the Royal Astronomical Society: Letters}\ }\textbf
  {\bibinfo {volume} {395}},\ \bibinfo {pages} {L48} (\bibinfo {year}
  {2009})}\BibitemShut {NoStop}%
\bibitem [{\citenamefont {Cameron}\ and\ \citenamefont
  {Alexakis}(2016)}]{cameron2016fate}%
  \BibitemOpen
  \bibfield  {author} {\bibinfo {author} {\bibfnamefont {A.}~\bibnamefont
  {Cameron}}\ and\ \bibinfo {author} {\bibfnamefont {A.}~\bibnamefont
  {Alexakis}},\ }\href@noop {} {\bibfield  {journal} {\bibinfo  {journal}
  {Physical review letters}\ }\textbf {\bibinfo {volume} {117}},\ \bibinfo
  {pages} {205101} (\bibinfo {year} {2016})}\BibitemShut {NoStop}%
\bibitem [{\citenamefont {Vainshtein}\ and\ \citenamefont
  {Cattaneo}(1992)}]{vainshtein1992nonlinear}%
  \BibitemOpen
  \bibfield  {author} {\bibinfo {author} {\bibfnamefont {S.~I.}\ \bibnamefont
  {Vainshtein}}\ and\ \bibinfo {author} {\bibfnamefont {F.}~\bibnamefont
  {Cattaneo}},\ }\href@noop {} {\bibfield  {journal} {\bibinfo  {journal} {The
  Astrophysical Journal}\ }\textbf {\bibinfo {volume} {393}},\ \bibinfo {pages}
  {165} (\bibinfo {year} {1992})}\BibitemShut {NoStop}%
\bibitem [{\citenamefont {Hughes}(2008)}]{hughes2008mean}%
  \BibitemOpen
  \bibfield  {author} {\bibinfo {author} {\bibfnamefont {D.}~\bibnamefont
  {Hughes}},\ }\href@noop {} {\bibfield  {journal} {\bibinfo  {journal} {Plasma
  Physics and Controlled Fusion}\ }\textbf {\bibinfo {volume} {50}},\ \bibinfo
  {pages} {124021} (\bibinfo {year} {2008})}\BibitemShut {NoStop}%
\bibitem [{\citenamefont {Blackman}\ and\ \citenamefont
  {Field}(2002)}]{blackman2002new}%
  \BibitemOpen
  \bibfield  {author} {\bibinfo {author} {\bibfnamefont {E.~G.}\ \bibnamefont
  {Blackman}}\ and\ \bibinfo {author} {\bibfnamefont {G.~B.}\ \bibnamefont
  {Field}},\ }\href@noop {} {\bibfield  {journal} {\bibinfo  {journal}
  {Physical Review Letters}\ }\textbf {\bibinfo {volume} {89}},\ \bibinfo
  {pages} {265007} (\bibinfo {year} {2002})}\BibitemShut {NoStop}%
\bibitem [{\citenamefont {Blackman}(2003)}]{blackman2003recent}%
  \BibitemOpen
  \bibfield  {author} {\bibinfo {author} {\bibfnamefont {E.~G.}\ \bibnamefont
  {Blackman}},\ }\href@noop {} {\bibfield  {journal} {\bibinfo  {journal}
  {Turbulence and Magnetic Fields in Astrophysics}\ ,\ \bibinfo {pages} {432}}
  (\bibinfo {year} {2003})}\BibitemShut {NoStop}%
\bibitem [{\citenamefont {Frisch}\ \emph {et~al.}(1975)\citenamefont {Frisch},
  \citenamefont {Pouquet}, \citenamefont {L{\'e}orat},\ and\ \citenamefont
  {Mazure}}]{frisch1975possibility}%
  \BibitemOpen
  \bibfield  {author} {\bibinfo {author} {\bibfnamefont {U.}~\bibnamefont
  {Frisch}}, \bibinfo {author} {\bibfnamefont {A.}~\bibnamefont {Pouquet}},
  \bibinfo {author} {\bibfnamefont {J.}~\bibnamefont {L{\'e}orat}}, \ and\
  \bibinfo {author} {\bibfnamefont {A.}~\bibnamefont {Mazure}},\ }\href@noop {}
  {\bibfield  {journal} {\bibinfo  {journal} {Journal of Fluid Mechanics}\
  }\textbf {\bibinfo {volume} {68}},\ \bibinfo {pages} {769} (\bibinfo {year}
  {1975})}\BibitemShut {NoStop}%
\bibitem [{\citenamefont {Pouquet}\ \emph {et~al.}(1976)\citenamefont
  {Pouquet}, \citenamefont {Frisch},\ and\ \citenamefont
  {L{\'e}orat}}]{pouquet1976strong}%
  \BibitemOpen
  \bibfield  {author} {\bibinfo {author} {\bibfnamefont {A.}~\bibnamefont
  {Pouquet}}, \bibinfo {author} {\bibfnamefont {U.}~\bibnamefont {Frisch}}, \
  and\ \bibinfo {author} {\bibfnamefont {J.}~\bibnamefont {L{\'e}orat}},\
  }\href@noop {} {\bibfield  {journal} {\bibinfo  {journal} {Journal of Fluid
  Mechanics}\ }\textbf {\bibinfo {volume} {77}},\ \bibinfo {pages} {321}
  (\bibinfo {year} {1976})}\BibitemShut {NoStop}%
\bibitem [{\citenamefont {Pouquet}\ and\ \citenamefont
  {Patterson}(1978)}]{pouquet1978numerical}%
  \BibitemOpen
  \bibfield  {author} {\bibinfo {author} {\bibfnamefont {A.}~\bibnamefont
  {Pouquet}}\ and\ \bibinfo {author} {\bibfnamefont {G.}~\bibnamefont
  {Patterson}},\ }\href@noop {} {\bibfield  {journal} {\bibinfo  {journal}
  {Journal of Fluid Mechanics}\ }\textbf {\bibinfo {volume} {85}},\ \bibinfo
  {pages} {305} (\bibinfo {year} {1978})}\BibitemShut {NoStop}%
\bibitem [{\citenamefont {Brandenburg}(2001)}]{brandenburg2001inverse}%
  \BibitemOpen
  \bibfield  {author} {\bibinfo {author} {\bibfnamefont {A.}~\bibnamefont
  {Brandenburg}},\ }\href@noop {} {\bibfield  {journal} {\bibinfo  {journal}
  {The Astrophysical Journal}\ }\textbf {\bibinfo {volume} {550}},\ \bibinfo
  {pages} {824} (\bibinfo {year} {2001})}\BibitemShut {NoStop}%
\bibitem [{\citenamefont {Alexakis}\ \emph {et~al.}(2006)\citenamefont
  {Alexakis}, \citenamefont {Mininni},\ and\ \citenamefont
  {Pouquet}}]{alexakis2006inverse}%
  \BibitemOpen
  \bibfield  {author} {\bibinfo {author} {\bibfnamefont {A.}~\bibnamefont
  {Alexakis}}, \bibinfo {author} {\bibfnamefont {P.~D.}\ \bibnamefont
  {Mininni}}, \ and\ \bibinfo {author} {\bibfnamefont {A.}~\bibnamefont
  {Pouquet}},\ }\href@noop {} {\bibfield  {journal} {\bibinfo  {journal} {The
  Astrophysical Journal}\ }\textbf {\bibinfo {volume} {640}},\ \bibinfo {pages}
  {335} (\bibinfo {year} {2006})}\BibitemShut {NoStop}%
\bibitem [{\citenamefont {M{\"u}ller}\ \emph {et~al.}(2012)\citenamefont
  {M{\"u}ller}, \citenamefont {Malapaka},\ and\ \citenamefont
  {Busse}}]{muller2012inverse}%
  \BibitemOpen
  \bibfield  {author} {\bibinfo {author} {\bibfnamefont {W.-C.}\ \bibnamefont
  {M{\"u}ller}}, \bibinfo {author} {\bibfnamefont {S.~K.}\ \bibnamefont
  {Malapaka}}, \ and\ \bibinfo {author} {\bibfnamefont {A.}~\bibnamefont
  {Busse}},\ }\href@noop {} {\bibfield  {journal} {\bibinfo  {journal}
  {Physical Review E}\ }\textbf {\bibinfo {volume} {85}},\ \bibinfo {pages}
  {015302} (\bibinfo {year} {2012})}\BibitemShut {NoStop}%
\bibitem [{\citenamefont {Bhat}\ \emph {et~al.}(2019)\citenamefont {Bhat},
  \citenamefont {Subramanian},\ and\ \citenamefont
  {Brandenburg}}]{bhat2019efficient}%
  \BibitemOpen
  \bibfield  {author} {\bibinfo {author} {\bibfnamefont {P.}~\bibnamefont
  {Bhat}}, \bibinfo {author} {\bibfnamefont {K.}~\bibnamefont {Subramanian}}, \
  and\ \bibinfo {author} {\bibfnamefont {A.}~\bibnamefont {Brandenburg}},\
  }\href@noop {} {\bibfield  {journal} {\bibinfo  {journal} {arXiv preprint
  arXiv:1905.08278}\ } (\bibinfo {year} {2019})}\BibitemShut {NoStop}%
\bibitem [{\citenamefont {Schekochihin}\ \emph {et~al.}(2002)\citenamefont
  {Schekochihin}, \citenamefont {Boldyrev},\ and\ \citenamefont
  {Kulsrud}}]{schekochihin2002spectra}%
  \BibitemOpen
  \bibfield  {author} {\bibinfo {author} {\bibfnamefont {A.~A.}\ \bibnamefont
  {Schekochihin}}, \bibinfo {author} {\bibfnamefont {S.~A.}\ \bibnamefont
  {Boldyrev}}, \ and\ \bibinfo {author} {\bibfnamefont {R.~M.}\ \bibnamefont
  {Kulsrud}},\ }\href@noop {} {\bibfield  {journal} {\bibinfo  {journal} {The
  Astrophysical Journal}\ }\textbf {\bibinfo {volume} {567}},\ \bibinfo {pages}
  {828} (\bibinfo {year} {2002})}\BibitemShut {NoStop}%
\bibitem [{\citenamefont {Biskamp}(2003)}]{biskamp2003magnetohydrodynamic}%
  \BibitemOpen
  \bibfield  {author} {\bibinfo {author} {\bibfnamefont {D.}~\bibnamefont
  {Biskamp}},\ }\href@noop {} {\emph {\bibinfo {title} {Magnetohydrodynamic
  turbulence}}}\ (\bibinfo  {publisher} {Cambridge University Press},\ \bibinfo
  {year} {2003})\BibitemShut {NoStop}%
\bibitem [{\citenamefont {Alexakis}\ and\ \citenamefont
  {Biferale}(2018)}]{alexakis2018cascades}%
  \BibitemOpen
  \bibfield  {author} {\bibinfo {author} {\bibfnamefont {A.}~\bibnamefont
  {Alexakis}}\ and\ \bibinfo {author} {\bibfnamefont {L.}~\bibnamefont
  {Biferale}},\ }\href@noop {} {\bibfield  {journal} {\bibinfo  {journal}
  {Physics Reports}\ }\textbf {\bibinfo {volume} {767}},\ \bibinfo {pages} {1}
  (\bibinfo {year} {2018})}\BibitemShut {NoStop}%
\bibitem [{\citenamefont {Waleffe}(1992)}]{waleffe1992nature}%
  \BibitemOpen
  \bibfield  {author} {\bibinfo {author} {\bibfnamefont {F.}~\bibnamefont
  {Waleffe}},\ }\href@noop {} {\bibfield  {journal} {\bibinfo  {journal}
  {Physics of Fluids A: Fluid Dynamics}\ }\textbf {\bibinfo {volume} {4}},\
  \bibinfo {pages} {350} (\bibinfo {year} {1992})}\BibitemShut {NoStop}%
\bibitem [{\citenamefont {Linkmann}\ \emph {et~al.}(2016)\citenamefont
  {Linkmann}, \citenamefont {Berera}, \citenamefont {McKay},\ and\
  \citenamefont {J{\"a}ger}}]{linkmann2016helical}%
  \BibitemOpen
  \bibfield  {author} {\bibinfo {author} {\bibfnamefont {M.}~\bibnamefont
  {Linkmann}}, \bibinfo {author} {\bibfnamefont {A.}~\bibnamefont {Berera}},
  \bibinfo {author} {\bibfnamefont {M.}~\bibnamefont {McKay}}, \ and\ \bibinfo
  {author} {\bibfnamefont {J.}~\bibnamefont {J{\"a}ger}},\ }\href@noop {}
  {\bibfield  {journal} {\bibinfo  {journal} {Journal of Fluid Mechanics}\
  }\textbf {\bibinfo {volume} {791}},\ \bibinfo {pages} {61} (\bibinfo {year}
  {2016})}\BibitemShut {NoStop}%
\bibitem [{\citenamefont {Rincon}(2021)}]{rincon2021helical}%
  \BibitemOpen
  \bibfield  {author} {\bibinfo {author} {\bibfnamefont {F.}~\bibnamefont
  {Rincon}},\ }\href@noop {} {\bibfield  {journal} {\bibinfo  {journal}
  {Physical Review Fluids}\ }\textbf {\bibinfo {volume} {6}},\ \bibinfo {pages}
  {L121701} (\bibinfo {year} {2021})}\BibitemShut {NoStop}%
\bibitem [{\citenamefont {Cattaneo}\ \emph {et~al.}(2020)\citenamefont
  {Cattaneo}, \citenamefont {Bodo},\ and\ \citenamefont
  {Tobias}}]{cattaneo2020magnetic}%
  \BibitemOpen
  \bibfield  {author} {\bibinfo {author} {\bibfnamefont {F.}~\bibnamefont
  {Cattaneo}}, \bibinfo {author} {\bibfnamefont {G.}~\bibnamefont {Bodo}}, \
  and\ \bibinfo {author} {\bibfnamefont {S.}~\bibnamefont {Tobias}},\
  }\href@noop {} {\bibfield  {journal} {\bibinfo  {journal} {Journal of Plasma
  Physics}\ }\textbf {\bibinfo {volume} {86}} (\bibinfo {year}
  {2020})}\BibitemShut {NoStop}%
\bibitem [{\citenamefont {Galtier}\ \emph {et~al.}(2000)\citenamefont
  {Galtier}, \citenamefont {Nazarenko}, \citenamefont {Newell},\ and\
  \citenamefont {Pouquet}}]{galtier2000weak}%
  \BibitemOpen
  \bibfield  {author} {\bibinfo {author} {\bibfnamefont {S.}~\bibnamefont
  {Galtier}}, \bibinfo {author} {\bibfnamefont {S.}~\bibnamefont {Nazarenko}},
  \bibinfo {author} {\bibfnamefont {A.~C.}\ \bibnamefont {Newell}}, \ and\
  \bibinfo {author} {\bibfnamefont {A.}~\bibnamefont {Pouquet}},\ }\href@noop
  {} {\bibfield  {journal} {\bibinfo  {journal} {Journal of plasma physics}\
  }\textbf {\bibinfo {volume} {63}},\ \bibinfo {pages} {447} (\bibinfo {year}
  {2000})}\BibitemShut {NoStop}%
\bibitem [{\citenamefont {Alexakis}(2013)}]{alexakis2013large}%
  \BibitemOpen
  \bibfield  {author} {\bibinfo {author} {\bibfnamefont {A.}~\bibnamefont
  {Alexakis}},\ }\href@noop {} {\bibfield  {journal} {\bibinfo  {journal}
  {Physical Review Letters}\ }\textbf {\bibinfo {volume} {110}},\ \bibinfo
  {pages} {084502} (\bibinfo {year} {2013})}\BibitemShut {NoStop}%
\bibitem [{\citenamefont {Mininni}\ \emph {et~al.}(2011)\citenamefont
  {Mininni}, \citenamefont {Rosenberg}, \citenamefont {Reddy},\ and\
  \citenamefont {Pouquet}}]{mininni2011hybrid}%
  \BibitemOpen
  \bibfield  {author} {\bibinfo {author} {\bibfnamefont {P.~D.}\ \bibnamefont
  {Mininni}}, \bibinfo {author} {\bibfnamefont {D.}~\bibnamefont {Rosenberg}},
  \bibinfo {author} {\bibfnamefont {R.}~\bibnamefont {Reddy}}, \ and\ \bibinfo
  {author} {\bibfnamefont {A.}~\bibnamefont {Pouquet}},\ }\href@noop {}
  {\bibfield  {journal} {\bibinfo  {journal} {Parallel computing}\ }\textbf
  {\bibinfo {volume} {37}},\ \bibinfo {pages} {316} (\bibinfo {year}
  {2011})}\BibitemShut {NoStop}%
\end{thebibliography}%

\end{document}